\documentstyle[11pt,newpasp,twoside,epsf]{article}
\def\edcomment#1{\iffalse\marginpar{\raggedright\sl#1\/}\else\relax\fi}
\reversemarginpar

\newcommand{\approxgt}{\mbox{$\;^{>}\hspace{-0.24cm}_{\sim}\;$}}

\begin{document}
\title{Radio Polarimetry Results for Young Southern Pulsars}
\author{Fronefield Crawford, Victoria M. Kaspi}
\affil{Center for Space Research, MIT, Cambridge, MA 02139, USA}
\author{Richard N. Manchester}
\affil{ATNF, CSIRO, P.O. Box 76, Epping, NSW, Australia}

\begin{abstract}
We present radio polarimetry results for nine Southern pulsars. Six of
the nine are young, with characteristic ages less than 100 kyr and
high spin-down luminosities. All six show significant linear
polarization, and we confirm a previously noticed trend in which the
degree of linear polarization increases with spin-down luminosity.  We
have used the rotating vector model to fit the observed position angle
data for PSR J1513$-$5908 (B1509$-$58). We find that a magnetic
inclination angle $\alpha > 60^{\circ}$ is excluded at the 3$\sigma$
level in the fit, and that the geometry suggested by the morphology of
an apparent bipolar X-ray outflow is marginally inconsistent with a
recent model of the pulsar magnetosphere.
\end{abstract}

\noindent We present polarimetry results for nine Southern pulsars,
six of which are young ($\tau_{c} < 100$ kyr) with high spin-down
luminosities ($\dot{E} > 10^{34}$ erg/s). Our results complete the
polarization information for the population of currently known young
pulsars which are bright enough to be detected in radio polarization.
All of the pulsars in our sample were observed at 1350 MHz with the
Parkes 64-meter radio telescope, and in several cases were also
observed at 660 MHz and 2260 MHz. The hardware setup and observing
technique are the same as those described elsewhere (Navarro 1994;
Manchester, Han, \& Qiao 1998).

All six young pulsars show significant linear polarization at 1350 MHz
(PSRs J1105$-$6107, J1341$-$6220, J1513$-$5908, J1646$-$4346,
J1730$-$3350, and J1801$-$2306).  We confirm a previously noticed
trend in which the degree of linear polarization is positively
correlated with spin-down luminosity. This trend has been shown to be
stronger at 4.9 GHz (von Hoensbroech, Kijak, \& Krawczyk 1998) and is
not evident at 400 MHz (Gould \& Lyne 1998). Our results, which are
concentrated at the high end of the $\dot{E}$ range in Fig. 1a, are
consistent with this trend at 1350 MHz.

We have fit the rotating vector model of the position angle (PA) swing
(Radhakrishnan \& Cooke 1969) to the 1350 MHz data on PSR J1513$-$5908
(B1509$-$58). The PA swing is flat, consistent with emission coming
from a partial conal beam (Manchester 1996). The polarization profile
and PA swing are shown in Fig.~1b. From our PA fit we find a
magnetic inclination angle range of $\alpha < 60^{\circ}$ (3$\sigma$
confidence). Imposing the restriction that the angle between the spin
axis and observer's line of sight is $\zeta \approxgt 70^{\circ}$, as
proposed by Brazier \& Becker (1997) using the morphology of an
apparent bipolar X-ray outflow, implies $\alpha > 30^{\circ}$ in our
fit. In a model proposed by Melatos (1997) in which $P$, $\dot{P}$,
and $\alpha$ uniquely determine the braking index $n$ of a pulsar, the
range $\alpha > 30^{\circ}$ for PSR J1513$-$5908 implies $n >$
2.86. This is marginally inconsistent with the observed value of $n =
2.837 \pm 0.001$ (Kaspi et al. 1994). The result from our fit
indicates that the Melatos model and the geometry $\zeta \approxgt
70^{\circ}$ implied by Brazier \& Becker may be inconsistent with each
other. The Melatos model may also be more directly tested using PSR
J1513$-$5908 with the restriction $\alpha < 60^{\circ}$ in combination
with refinements in the spin parameters from future timing
observations. We have submitted an article to the Astronomical Journal
which details the results presented here.

\begin{figure}
\epsfysize=2.70in
\centerline{\epsfbox{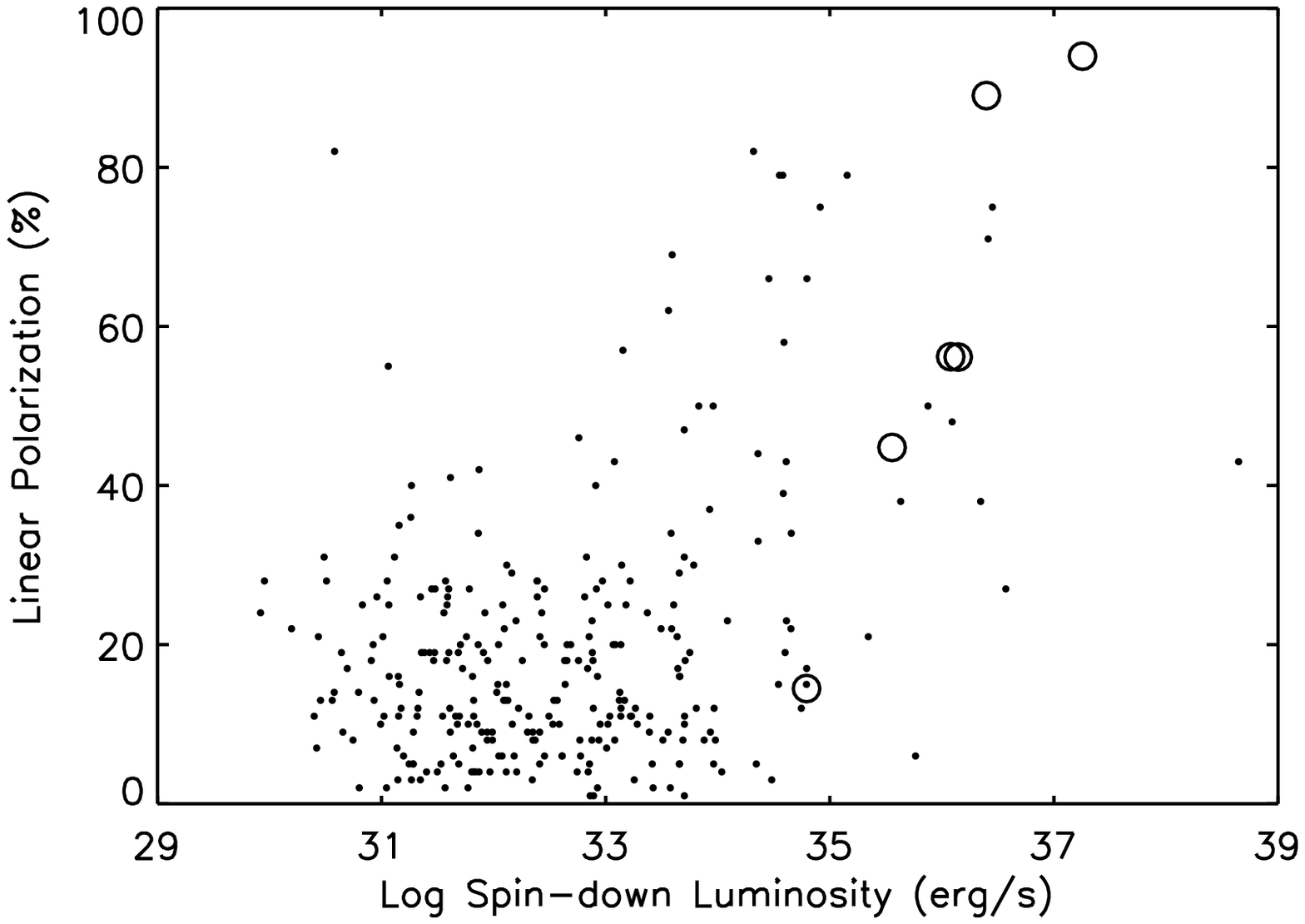} \epsfysize=3.00in \epsfbox{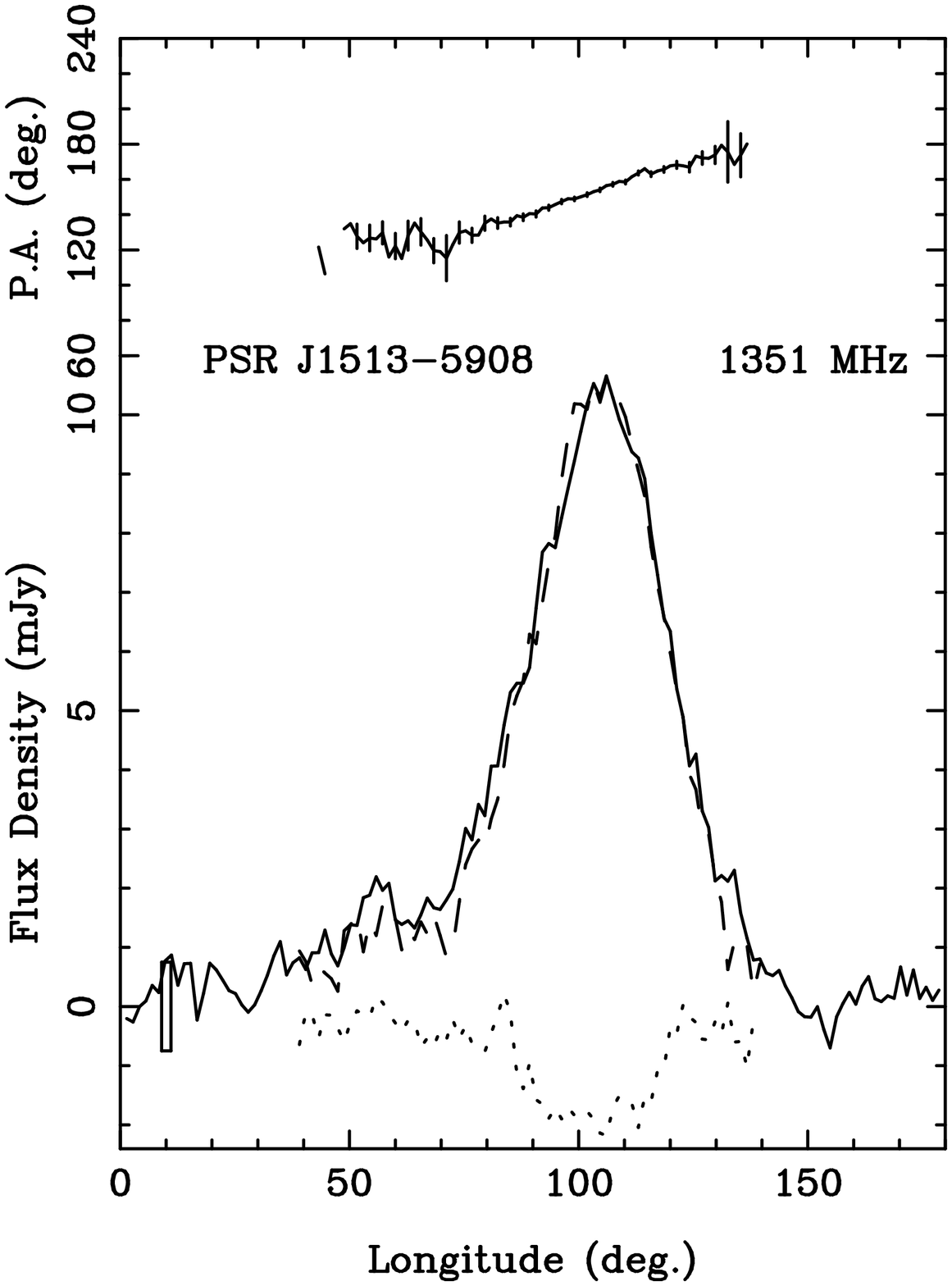}}
\caption{Left: Fractional linear polarization as a function of log
$\dot{E}$ for the six young pulsars in our sample at 1350 MHz (open
circles) and for 278 pulsars from Gould \& Lyne (1998) at 1400 MHz
(dots). Right: 1350 MHz polarization profile for PSR J1513$-$5908 with
position angle variation on top. The pulsar is almost completely
linearly polarized.}
\end{figure}

\references 

Brazier, K. T. S. \& Becker, W. 1997, \mnras, 284, 335 \\
Gould, D. M. \& Lyne, A. G. 1998, \mnras, 301, 235 \\
Kaspi, V. M., Manchester, R. N., Siegman, B., Johnston, S., \& Lyne,
A. G. 1994, \apjl, 409, L57 \\
Manchester, R. N. 1996, IAU Colloq. 160, {\it Pulsars: Problems and Progress}, 
193 \\ 
Manchester, R. N., Han, J. L., \& Qiao, G. J. 1998, \mnras, 295, 280 \\
Melatos, A. 1997, \mnras, 288, 1049 \\
Navarro, J. 1994, Ph.D. Thesis, California Institute of Technology \\
Radhakrishnan, V. \& Cooke, D. J. 1969, Astrophys. Lett., 3, 225 \\
von Hoensbroech, A., Kijak, J., \& Krawczyk, A. 1998, \aap, 334, 571 

\end{document}